\documentstyle[12pt]{article}
\newcommand{\beq}{\begin{equation}}
\newcommand{\eeq}{\end{equation}}
\newcommand{\bea}{\begin{eqnarray}}
\newcommand{\eea}{\end{eqnarray}}

\begin{document}
%%%%%%%%%%%%%%%%%%%%%%%%%%%%%%%%%%%%%%%%%%%%%%%%%%%%%%%%%%%%%%%%%%%%%%
\title          {Tunneling between fermionic vacua 
                    and the overlap formalism.}
\author{
\rule{0cm}{1.cm}
C.D. Fosco\thanks{Member of  CONICET, Argentina.}
\\
{\normalsize\it 
Centro At\'omico Bariloche}
\\
{\normalsize\it 
8400 S.C. de Bariloche, Argentina}
}
\maketitle
%===================================================================
\begin{abstract}
The probability amplitude for tunneling between the Dirac vacua
corresponding to different signs of a parity breaking fermionic mass 
$M$ in $2+1$ dimensions is studied, making contact with the continuum 
overlap formulation for chiral determinants. 
It is shown that the transition probability in the limit when 
$M \to \infty$ corresponds, via the overlap formalism, to the
squared modulus of a chiral determinant in two Euclidean
dimensions.
The transition probabilities 
corresponding to two particular examples:  fermions on a torus with 
twisted boundary conditions, and fermions on a disk in the presence
of an external constant magnetic field are evaluated.
\end{abstract}
\newpage
%===================================================================
\section{Introduction.}
Quantum mechanical tunneling between different classical vacua is a
phenomenon whose importance in Quantum Field Theory should not need
to be stressed. A paradigmatic example of this are instanton
effects and the related tunneling between vacua in Yang-Mills 
theories, essential to the understanding of the vacuum structure
of such models~\cite{raja}. 
No less relevant to tunneling effects is the role of fermions.
Under certain conditions their contribution to the transition
probability can alter dramatically the results obtained from
the other fields involved. This is particularly true of 
models with massless fermions couple to a gauge field with non-zero
Pontryagin index in $4$ dimensions, where the presence of fermionic
zero modes suppress the transition probability~\cite{cole}.

In this letter we study the probability amplitude for fermionic transition 
between vacua in $2+1$ dimensions, the different vacua being
characterized by different signs of the fermion mass. This situation is
concretely realised by introducing a real pseudoscalar field with a Yukawa
coupling to the fermions, and having a spontaneous symmetry breaking
potential with two minima, one for each sign of the mass.
The fermions are also coupled to an external gauge field $A_\mu$
with $A_0 = 0$ and $A_j = A_j ({\vec x})$. A question that arises
naturally is
which is the dependence of the transition probability on the
external gauge field configuration and the parameters of the pseudoscalar
field potential. 

The relevance of this problem to the understanding of Quantum Electrodynamics
in $2+1$ dimensions stems from the fact that different signs of the mass
correspond to different parity violating configurations. The existence of
fermionic tunneling between them should imply that the true vacuum is a
symmetric configuration of the two, with no net parity violation coming
from the fermionic sector. The possibility of having different signs for
the fermion mass does not necessarily come from the existence of a spontaneous
symmetry breaking potential. It may also come, for example, from the fact 
that one is dealing with the massless theory, whose infrared divergences render
it ill-defined. The mass could then be seen as an infrared regulator,
whose sign determines the properties of the vacuum.

When defining the probability amplitude, one is faced with an object
which has an striking similarity with the squared modulus of the
so-called `overlap', the basic construct in the 
overlap formalism~\cite{nar}. This is a  proposal, based on an 
earlier idea of Kaplan~\cite{kap}, to define fermionic chiral 
determinants.  When implemented on the lattice, it seems to 
overcome the kinematical constraint stated by the Nielsen-Ninomiya 
theorem~\cite{kn:Karsten.}, and thus could provide a suitable
framework to study non-perturbative phenomena in models 
containing chiral fermions. 

In this  method, the determinant of a {\em massless \/} chiral Dirac operator in 
$2 d$ dimensions is defined as an overlap between the
Dirac vacuum states of two auxiliary Hamiltonians  acting on
{\em massive \/} Dirac fermions in $2 d + 1$ dimensions.  This 
implies that the probability amplitude for tunneling between vacua 
corresponds,
when the modulus of $M$ tends to infinity, to the squared modulus of 
a chiral determinant in $2 d$ dimensions. The presence of the modulus
guarantees the gauge invariance of the results.

This paper is organized as follows: In section 2 the kind of model
we are studying is introduced and the corresponding transition probability
defined. The formula for the transition probability is evaluated
for the case of twisted boundary conditions on a torus in section 3,
and for a constant magnetic field in section 4. In section 5 we
discuss our results and present our conclusions.

\newpage
\section{Transition probability.}
We shall consider a fermionic field interacting with an external
(i.e., non-dynamical) gauge field $A_\mu$ and a dynamical
pseudoscalar field $\varphi$ in $2+1$ dimensions. 
The system is described by the action
$$ S \; =\; \int \; d^3 x \; {\cal L} \;\;\;\;\;\;\;\; 
   {\cal L} \;=\; {\cal L}_F \, + \, {\cal L}_{\varphi} $$
\beq
{\cal L}_F \;=\; {\bar \psi} ( i \not \! \partial 
- e \not \! A  - g \varphi ) \psi \;\;\;\;\;\;
{\cal L}_{\varphi} \;=\; \frac{1}{2} (\partial \varphi)^2 \,
- {\displaystyle \frac{\lambda}{2}} (\varphi^2 - \varphi_0^2)^2 
\label{action}
\eeq
where $\lambda > 0$. The form we have chosen for the pseudoscalar
field Lagrangian is such that spontaneous symmetry breaking
does occur, the two classical vacua for this field being
simply $\varphi = \pm \varphi_0$. To study the vacuum structure 
of the full system, we  need to put the fermions in their vacua 
as well.
As the fermionic Lagrangian  is sensitive to the sign chosen for 
the vacuum value of $\varphi$, the form it adopts for each of the
$\varphi$ vacua has to be studied separately. This yields  
\beq
{\cal L}_F^{\pm} \;=\; {\bar \psi} ( i \not \! \partial
- e \not \!\! A \mp M ) \psi 
\label{lpm}
\eeq
where $M = g \varphi_0$. The fermionic vacua will simply
be the two possible Dirac vacua corresponding to ${\cal L}_F^+$ and
${\cal L}_F^-$. We shall deal with gauge field configurations
such that $A_0 = 0$ and $A_j$ is static (in order to have
a vacuum state). 
Dirac vacua are constructed by filling all the 
negative energy states. The second-quantized Hamiltonians corresponding 
to (\ref{lpm}):
\beq
H_\pm (A) \;=\; \int d^2 x \, \Psi^{\dagger} (x) 
{\cal H}_\pm (A) \Psi (x) 
\label{sqh}
\eeq
where $\Psi (x)$ is the fermionic field operator, and 
${\cal H}_\pm$ are the two 1-body Dirac Hamiltonians
\beq
{\cal H}_\pm (A) \;=\; {\vec \alpha} \cdot ( -i {\vec \partial}
- e {\vec A} ) \pm \beta M \;.
\label{fqh}
\eeq

We define the eigenstates of ${\cal H}_\pm
(A)$
\bea 
{\cal H}_\pm (A) \; u_\pm (\lambda | A, x) &=& 
\omega (\lambda | A, x)\; u_\pm (\lambda | A, x) \nonumber\\
{\cal H}_\pm (A) \; v_\pm (\lambda | A, x) &=& 
- \omega (\lambda | A, x) \; v_\pm (\lambda | A, x)
\label{eigen} 
\eea
where $u_\pm (\lambda | A, x)$ and $v_\pm (\lambda | A, x)$ are
the positive and negative energy eigenstates, respectively.
$\lambda$ is an index which labels the eigenstates, and is
assumed to be discrete for the sake of simplicity, but a continuous
spectrum can also be dealt with in a completely analogous manner. 
The $x$ and $A$ dependence of the
eigenspinors, as well as the $A$-dependence of the energy 
$\omega$ have been made explicit.
Note that the above objects do also depend on the value of the mass $M$.
We assume the orthonormality relations
$$ (\;\; u_\pm (\lambda | A, x) , u_\pm (\lambda' | A, x) \;\;)\;=\; 
\delta_{\lambda,\lambda'}
\;\;\;\;
(\;\; v_\pm (\lambda | A, x) , v_\pm (\lambda' | A, x) \;\;) \;=\; \delta_{\lambda,\lambda'}$$
\beq
(\;\; v_\pm (\lambda | A, x) , u_\pm (\lambda' | A, x) \;\; )\;=\; 0 
\label{orth}
\eeq
where the scalar products mean inner product in spinorial
two-component space as well as integration over the coordinates.
In terms of the eigenspinors (\ref{eigen}), the fermionic field may be expanded
using either sign of the mass,
\bea
\Psi (x) &=& \sum_{\lambda} \left[ \;\; b_\pm (\lambda) \; u_\pm (\lambda | A, x)
+ d^\dagger_\pm (\lambda)  \; v_\pm (\lambda | A, x) \;\; \right] \nonumber\\
\Psi^\dagger (x) &=& \sum_{\lambda} \left[ \;\; b^\dagger_\pm (\lambda) 
\; u^\dagger_\pm 
(\lambda | A, x) + d_\pm (\lambda)\; v^\dagger_\pm (\lambda | A, x) \;\;\right]
\; . 
\label{exp}
\eea
Whence the canonical anticommutation relations become
\beq
\{ \Psi_\alpha (x) \; , \; \Psi_\beta (y) \} = 0 \;\;\;\;\;\;
\{ \Psi_\alpha (x) \; , \; \Psi_{\beta}^{\dagger} (y) \} = \delta_{\alpha \beta} 
\, \delta^{(2)} (x-y) \; .
\label{acr}
\eeq
By using the expansions defined by Eqs.(\ref{eigen}-\ref{exp}), the Dirac vacua
corresponding to the two possible signs of the mass become
\beq
| A \; \pm > \;\;\; = \;\;\; \left[ \prod_\lambda 
d_\pm (\lambda) \right] \; | 0 > 
\label{vacua}
\eeq
where $|0>$ is the Fock vacuum, which satisfies $\Psi (x) |0> = 0 \, , \forall x$,
and the product runs over all the possible values of the index $\lambda$.
It goes without saying that the transition probability, normalized to $1$
for $A_\mu = 0$ is  
\beq
\displaystyle{\frac{P_\pm (A)}{P_\pm (0)}} \;=\; | 
\displaystyle{\frac{<A + | A ->}{< + | - >}} |^2
\label{over}
\eeq
where $|\pm > \equiv |0 \pm >$. In ({\ref{over}) we recognize the
object which appears in the overlap definition of the chiral
determinant, in this case in two Euclidean dimensions.
More precisely, in the overlap formalism one can obtain
the squared modulus of the normalized chiral determinant by means
of the limit~\cite{overl}
\beq
| \frac{\det [ (\not \! \partial + i e \not \! A ) 
(\displaystyle{\frac{1 + \gamma_5}{2}}) ]}{\det [ \not \! \partial 
(\displaystyle{\frac{1 + \gamma_5}{2}}) ]}|^2
\;=\; \lim_{M \to \infty} \; | \displaystyle{\frac{<A + | A ->}{< + | - >}} |^2
\label{defover}
\eeq
where the operators in (\ref{defover}) act on functions defined on two-dimensional
Euclidean space. For example
\beq
\not \! \partial \;=\; \gamma_1 \partial_1 + \gamma_2 \partial_2 \; ,
\; \gamma_1 = \sigma_1 \; , \; \gamma_2 = \sigma_2 \; , \; 
\; \gamma_3 = \sigma_3 \; .
\eeq
The overlap between the two vacua may be given in terms of the negative
energy eigenstates (\ref{eigen}) of ${\cal H}_\pm (A)$. We first use
(\ref{exp}) to write
\beq
< A \; + | A \; - > \; = \;  < 0 | [ \prod_\lambda d_+ (\lambda) ]^{\dagger}
\; [ \prod_{\lambda'} d_- (\lambda') ] | 0 > \; .
\label{prodop} 
\eeq
We then  use the relation between the creation and annihilation operators
defined for both signs of the mass $M$
\bea
b_+ (\lambda) &=& 
\sum_{\lambda'} 
\left[ (u_+ (\lambda) , u_- (\lambda')) b_- (\lambda') \; + \;
(u_+ (\lambda) , v_- (\lambda')) d_-^{\dagger} (\lambda') \right] \nonumber\\
d_+^{\dagger} (\lambda) &=& 
\sum_{\lambda'} 
\left[ (v_+ (\lambda) , u_- (\lambda')) b_- (\lambda') \; + \;
(v_+ (\lambda) , v_- (\lambda')) d_-^{\dagger} (\lambda') \right]\;.
\label{bogo}
\eea
By using the second line of (\ref{bogo}) into (\ref{prodop}), we may write
each factor $d_+^{\dagger}$ in terms of the full set of operators
$b_-(\lambda)$ and $d_-(\lambda)$.  
The terms proportional to $b_- (\lambda)$ do not contribute because they can be 
anticommuted with all the $d_- (\lambda)$ operators, and they finally
annihilate the Fock vacuum. Thus only the contributions depending on
the scalar product between $v_+ (\lambda)$ and $v_- (\lambda)$ survive,
yielding
\beq
< A \; + | A \; - > \; = \; \det_{\lambda , \lambda'} (\;\; v_+ (\lambda) ,
v_- (\lambda') \;\;) \; . 
\label{prodpr}
\eeq
Hence the ratio between probabilities given in (\ref{over}) yields
\beq
\displaystyle{\frac{P_\pm (A)}{P_\pm (0)}} \;=\;
| \frac{\det_{\lambda , \lambda'} (v_+ (A| \lambda) ,
v_- (A | \lambda') )}{\det_{\lambda , \lambda'} (v_+ (0 | \lambda) ,
v_- (0| \lambda') )} |^2
\; .
\label{prodnorm}
\eeq

We shall now evaluate (\ref{prodnorm}) in two situations of interest:
(1) Fermions on a torus with constant metric and arbitrary twisting ,
(2) Fermions in the presence of an external magnetic field. 

%=================================================================
\section{Fermions with twisted boundary conditions on a torus.}
The two-dimensional torus is coordinatized by two real
variables 
\begin{equation}
\sigma^1 \;\;,\;\; \sigma^2 \;\;\;\; ,\;\;\;\; 0 \leq \sigma^{\mu} \leq 1 \;,
\label{1.1}
\end{equation}
and is equipped with the Euclidean metric
\begin{eqnarray}
ds^2 \,&=&\, \mid d \sigma^1 + \tau d \sigma^2 \mid^2 \,=\,
g_{\mu \nu} \; d\sigma^{\mu} d\sigma^{\nu}  \;,\nonumber\\
g_{\mu\nu} \,&=&\,
\left( \begin{array}{cc} 1 & \tau_1 \\ \tau_1 & \mid \tau \mid^2
\end{array} \right) \;\;\;\;\; g^{\mu\nu} \,=\,\frac{1}{\tau_2^2} \,
\left( \begin{array}{cc} \mid \tau \mid^2 & -\tau_1 \\ -\tau_1 & 1
\end{array} \right)
\label{1.2}
\end{eqnarray}
where $\tau = \tau_1 + i \tau_2$ and $\tau_2 > 0$.
 
The Dirac Hamiltonian operators then become
\begin{equation}
{\cal H}_\pm (A) \,=\, \sigma_3 ( \sigma^a \, e^{\mu}_a \, (\partial_{\mu} + i \,
A_{\mu}) \pm M ) \;,
\label{1.4}
\end{equation}
where the $\sigma^a$'s are the usual Pauli matrices
\begin{equation}
\sigma^1 \,=\, 
\left( \begin{array}{cc}
0 & 1 \\ 1 & 0 \\ \end{array} \right) \;\; , \;\;
\sigma^2 \,=\, 
\left( \begin{array}{cc}
0 & -i \\ i & 0 \\ \end{array}\right) \;\; , \;\;
\sigma^3 \,=\, 
\left( \begin{array}{cc}
1 & 0 \\ 0 & - 1 \\ \end{array}\right) \;.
\label{1.5}
\end{equation}
and $e^{\mu}_a$ are the zweibeins:
\begin{eqnarray} 
g_{\mu \nu} \;e^{\mu}_a \, e^{\nu}_b \,&=&\, \delta_{a b} \;,\nonumber\\
e^{\mu}_1 \,=\, ( \,1\,,\, 0 \, ) \;& & \; 
e^{\mu}_2 \,=\, (\,-\frac{\tau_1}{\tau_2} \,, \,\frac{1}{\tau_2} \,) \;\;. 
\label{1.3}
\end{eqnarray}
Note that $\tau_2$ 
is the volume of the torus
\beq
\tau_2 \;=\; \int d^2 \sigma \; \sqrt{ \det g_{\mu \nu} } \; . 
\eeq

We want to describe twisted fermions , i.e.,
the fermionic field has the boundary conditions
\begin{eqnarray}
\Psi ( \sigma^1 + 1 \,,\, \sigma^2 ) \,&=&\, - \, e^{2 \pi i
\varphi_1} \, \, \Psi (\sigma^1 \,,\, \sigma^2) \nonumber\\
\Psi ( \sigma^1 \,,\, \sigma^2 + 1) \,&=&\, - \, e^{2 \pi i
\varphi_2} \, \, \Psi (\sigma^1 \,,\, \sigma^2) \;\;,
\label{1.8}
\end{eqnarray}
where $\varphi_1$ and $\varphi_2$ are real numbers such that $0 \leq \varphi_{\mu}
\leq 1 $ and $n_1$, $n_2$ run over all the half-integers. 
The `reference' boundary condition about which we define the 
twistings is antiperiodicity in both directions, hence the minus signs
on the rhs in (\ref{1.8}). 

We take advantage of the calculation of the overlap
between the two Dirac vacua for this situation made in \cite{twist}.
It follows from \cite{twist} that the ratio between probabilities 
(\ref{prodnorm}) may be written as
\beq
\frac{P_\pm (\varphi )}{P_\pm (0)} \;=\;
\prod_{n_1,n_2} \left[ \frac{ (n + \varphi)^2 + \lambda^2}{(n + \varphi)^2} 
\; \frac{n^2}{(n^2 + \lambda^2)} \right] 
\eeq
where $\lambda = \frac{\tau_2 M}{2 \pi}$ and $n^{2} \equiv 
g_{\mu \nu} n^\mu n^\nu $. It is not possible to give an
exact analytic expression for (\ref{prodnorm}), valid for all possible
values of $M$. However, for $\lambda >> 1$, which means that the mass
$M$ should be large as compared with the momentum spacing, we 
can use the leading contribution coming from the
limit $\lambda \to \infty$, which is precisely one of the results given in
\cite{twist}.
Thus the probability ratio becomes equal to the squared modulus of the
chiral determinant in $2$ dimensions, which yields,
\beq
\frac{P_\pm (\varphi )}{P_\pm (0)} \;=\;
| \displaystyle{ \frac{\vartheta (\alpha , \tau)}{\vartheta (0,\tau)} }
|^2 \; e^{ - 2 \pi \tau_2 \varphi_1^2}
\label{res1}
\eeq
where $\alpha = \tau \varphi_1 - \varphi_2$ and 
\beq
\vartheta (\alpha , \tau) \; = \; \sum_{n = -\infty}^{n = \infty}
\, e^{ i \pi \tau n^2 \, + \, 2 \pi i n \alpha } \; .
\eeq
This corresponds to the squared
modulus of the corresponding chiral determinant, found by using conformal
field theory methods in ~\cite{alv}.

%=================================================================
\section{Constant magnetic field on a disc.}

In this case, we assume there is an external constant magnetic field
$B$, so that the gauge field verifies
\beq
e \, \epsilon_{j k} \partial_j A_k \; = \; B \;=\; {\rm const}
\eeq
which in the symmetric gauge is satisfied by the configuration
\beq
A_j \; = \; - \frac{B}{2 e} \, \epsilon_{j k} x_k
\eeq
and the Dirac Hamiltonians ${\cal H}_\pm (A)$ thus become
\beq 
\left(
\begin{array}{cc}
\pm M & 2 \partial + \displaystyle{\frac{B}{2}} {\bar z} \\
-2 {\bar \partial} + \displaystyle{\frac{B}{2}} z & \mp M 
\end{array}
\right)
\eeq
where $z \equiv x_1 + i x_2$ , ${\bar z} \equiv x_1 - i x_2$, and 
$\partial \equiv \frac{\partial}{\partial z}$.

The negative energy eigenstates (the only ones we need in order to
evaluate the transition probability) are given by
\beq
v_\pm (n,l) \,=\, \sqrt{\frac{\omega_n \pm M}{2 \omega_n}} \,
\left( 
\begin{array}{c}
-  \displaystyle{\frac{\sqrt{2 B}}{\omega_n \pm M}} \, {\hat a} f_{n,l} \\
    f_{n,l}
\end{array}
\right)
\eeq
where $\omega_n = \sqrt{ M^2 + s B n}$. $f_{n,l}$ are orthonormal
eigenfunctions satisfying
\beq
(-2 {\bar \partial} + \frac{B}{2} z)(2 \partial + \frac{B}{2} {\bar z})
f_{n,l} \;=\; 2 B n f_{n,l}
\eeq
and we define creation and annihilation operators ${\hat a }^\dagger$ and
${\hat a}$ by
\beq
{\hat a} \; = \; \displaystyle{\frac{ 2 \partial + \frac{B}{2} {\bar z} }{
{\sqrt{2 B}} } }
\;\;\;
{\hat a} \; = \; \displaystyle{\frac{ 2 \partial + \frac{B}{2} {\bar z} }{
{\sqrt{2 B}}}}
\; ,
\eeq
respectively.
The functions $f_{n,l}$ may be built by successive application of the creation 
operator ${\hat a}^{\dagger}$ on the $n = 0$ state
\bea
f_{n,l} &=& \displaystyle{ \frac{ ( {\hat a}^\dagger )^n }{\sqrt n !} } f_{0,l} 
\nonumber\\
f_{0,l} &=& [ \pi (\frac{2}{B})^{l+1} l! ]^{- \frac{1}{2}} 
e^{- \frac{B}{4} {\bar z} z } \, {\bar z}^l \;. 
\eea

The structure of the functions $f_{n,l}$ is quite similar to the ones one
encounters when studying the eigenstates of a non-relativistic particle
in the presence of an external magnetic field in two dimensions~\cite{frad}.
Indeed, in spite of the relativistic correction to the energies, the degeneracy
of every Landau level remains the same as in the non-relativistic case,
since this depends upon the structure of finite magnetic translations.
The degeneracy of each level is thus $d_n = N_\phi = \frac{B L^2}{2 \pi}$
where $N_\phi$ is the number of flux quanta piercing the disc, which is
assumed to be of finite size.

The result of evaluating (\ref{prodnorm}) can be put as
\beq
\frac{P (B) }{P (0)} \;=\; \prod_{n=1}^{\infty} 
(\frac{n}{n + x})^{N_\phi} 
\label{res2}
\eeq
where $x \equiv \frac{M^2}{2 B}$.
To evaluate (\ref{res2}), it is convenient to take logarithms on both sides 
\beq
\log P (B)  - \log P (0) \;=\; \frac{(M L)^2}{4 \pi} \; \sum_{n=1}^{\infty} 
x \; \log (\frac{n}{n + x}) \; .
\label{eqn}
\eeq
To evaluate (\ref{eqn}), we rewrite the logs by using Frullani's identity:
\beq
\log (\frac{a}{b}) \; = \; \lim_{\epsilon \to 0}
\, \int_{\epsilon}^{\infty} \, \frac{d s}{s} \, ( e^{- s b} \,-\, e^{- s a} )
\eeq
which, after some elementary manipulations yields
\beq
\frac{P(B)}{P(0)} \;=\; \exp \left[ \displaystyle{
\frac{(M L)^2}{2 \pi}} \, f(x) \right]
\eeq
where 
\beq
f (x) \;=\; \int_0^{\infty} \, \frac{d s }{s} \,
\displaystyle{\frac{ e^{-s x} - 1 + s x}{ e^s - 1}}\; .
\eeq   
%=================================================================
\section{Discussion.}
We have obtained the transition probabilities between the two
parity breaking vacua ($\pm M$), as functions of the 
parameters of the model for two particular gauge field configurations
(the twisted fermions can be equivalently regarded as untwisted but
in the presence of an external constant gauge field~\cite{twist}). 

It is worth remarking that there is a dependence of the 
probability ratio on the volume (area) of the system. 
In the case of the constant magnetic field, this dependence is such
that when $L \to \infty$, the ratio between probabilities tends
to zero. This is a manifestation of the fact that when the 
volume tends to infinity, there appears a zero mode for the
(two-dimensional) chiral determinant in the presence of a magnetic field.
For the case of twisted fermions on the torus, the dependence on 
the area is through the parameter $\tau_2$, which appears in a more
involved way, but it is not difficult to show that for large $\tau_2$ one
gets a probability ratio that is equal to zero when the 
components $\varphi$ are nearly half-integer. As in the constant magnetic 
field case, this is explained by the fact that the corresponding chiral
determinant develops a zero mode.

Finally, we note that there is a very important difference between the
transition probability one gets for the fermions and the corresponding to
the pseudoscalar field. For the fermions, the probability ratio
is {\em independent} of the parameter $\lambda$ which appears in the symmetry
breaking potential, and measures the height of the barrier. It only sees
the magnitude and sign of the vaccum value of $\varphi$.
Regarding the pseudoscalar field, if the volume is finite the 
transition probability may be non-zero, but it will vanish when
the height of the barrier diverges, since tunneling configurations will
cost an infinite action.

\newpage

\section*{Acknowledgements.}
I am grateful to S. Randjbar-Daemi , J. Strathdee and R.C. Trinchero for 
useful discussions.

\end{document}